\documentstyle[preprint,aps,eqsecnum]{revtex}
\begin{document}
\preprint{UCD 95-14, October 1995}
\draft
\title{Absence of physical walls in hot gauge theories}
\author{Joe Kiskis\thanks{email: jekiskis@ucdavis.edu  }}
\address{
Department of Physics\\
University of California, Davis, CA 95616, USA}
\date{\today}
\maketitle
\begin{abstract}
This paper shows that there are no {\em physical} walls
in the deconfined, high-temperature phase of $Z(2)$ lattice
gauge theory.
In a Hamiltonian formulation, the interface in the Wilson lines is not
physical. The line interface and its energy are interpreted in terms of
physical variables.
They are associated with a difference between two partition
functions. One includes only the configurations with even flux across the
interface. The other is restricted to odd flux.
Also, with matter present, there is no physical metastable state.
However, the free energy is lowered by the matter. This effect is
described in terms of physical variables.
\end{abstract}
\pacs{11.15.Ha, 5.50.+q, 75.10.Hk, 64.60.Ak}
\narrowtext

\section{Introduction}

This paper discusses some aspects of the global $Z(N)$ symmetry\cite{r1} of
finite-temperature $SU(N)$ or $Z(N)$ gauge theory.
It contributes to recent discussions of the physics of
$Z(N)$ phases and interfaces\cite{r2,me,r2.1,r2.2,r2.3,r2.4}.
The example of $Z(2)$ gauge theory is treated in detail here.
Before focusing on that case, a more general introduction is given.

There is a low-temperature phase of the theory, which is presumed to
confine fundamental sources, and a high-temperature phase, which does not.
Without matter fields, there is an order parameter---the
finite-temperature expectation value of the Wilson line $\langle L \rangle$.
The fundamental-representation
line $L(i)$ is the normalized trace of the
product of the group elements on the links in the inverse-temperature direction
at spatial site $i$.
\begin{equation}
 \langle L \rangle = \langle L(i) \rangle = \langle \case{1}{N}
                       Tr[\prod_{i_4} U(i,i_4,4)]\rangle
\end{equation}
The line carries a nontrivial representation of the global
$Z(N)$ symmetry. In the confining phase, $\langle L \rangle=0$, and the
ensemble
is $Z(N)$ symmetric. In the high $T$ phase, $\langle L \rangle$ takes one of
$N$ distinct values proportional to the $Nth$ roots of unity $z$ in $Z(N)$,
and the $Z(N)$ symmetry is broken.
The broken-symmetry, pure phases appear in Monte Carlo simulations, where runs
of finite duration in the
high-temperature phase and not too close to the critical temperature show the
Wilson line fluctuating around one of the $N$ available values.

In the Hamiltonian description\cite{r3},
the {\em physical} variables are the group elements on the links of the spatial
lattice.
In a Lagrangian formulation, there are also group
elements on links in the inverse-temperature direction.
These are unphysical, auxiliary variables introduced
to enforce the Gauss law constraints. The Wilson line is constructed from the
unphysical variables.
It is a projection operator that  forces the gauge field to be in a
fundamental rather than a singlet state at the spatial position of the line.
The global $Z(N)$ symmetry of the Lagrangian formulation is not physical; it
acts as the identity on all physical states\cite{me,r2.1,r2.2,r2.4}.
There is a single physical, high-temperature phase, which is the same for all
$z$.
In the Hamiltonian formulation, the high-temperature
phase is not distinguished by physical broken symmetry.
Rather the high-temperature
phase has a percolating flux network that is not present in the low-temperature
phase.

Now consider $Z(N)$ walls. If there were $N$, degenerate, physically-distinct,
high-temperature phases, then there could be large regions of space in
different phases and separated by interfaces. These walls could have a
physically-significant interface tension. In the line variables, there are
phase separations and interfaces. This topic has received much attention in the
literature\cite{r2,me,r2.1,r2.2,r2.3,r2.4}, and the interface tension for the
lines has been measured in Monte Carlo simulations\cite{r6}. However,  as
noted above, the $N$ phases of the Wilson lines are not physically distinct.
Thus one may question the physical relevance of the line interfaces.
There is work in the literature which is related this point and which asserts
that the interfaces are not physical\cite{r2.1,r2.2}.
There is considerable overlap of the present paper with some of that work.
Some of the statements in those papers are confirmed or derived by other
means here. Other issues treated here have not been previously addressed.

I use the
Hamiltonian description of finite-temperature gauge theory on a lattice in
three spatial dimensions in the $A_0=0$ gauge with Gauss law
constraints.  In
this formulation, the physical interpretation is most accessible.
I refer to this as the {\em physical} theory.
The {\em physical variables} are the group elements
on the links of the spatial lattice. They may be described with definite
position on the group or in a basis of definite irreducible representation
matrix elements, which is called the flux basis.
{\em Physical states} are configurations of physical variables
subject to Gauss law constraints. With that definition, the term ``physical''
takes on a specific, technical meaning. Alternative formulations of the quantum
gauge theory may lead to different pictures.

The constraints are enforced through the use of additional group
elements on sites. These are unphysical, auxiliary variables.
In a four-dimensional, Lagrangian formulation, they become the
elements on links in the inverse-temperature, 4-direction. The Wilson line
is constructed from them.

The interface exits in the line variables. It can be created and studied with
the application of appropriate boundary conditions. Applying the boundary
conditions and using physical variables gives a description in terms of flux
and reveals the property of flux that is being measured when the
line interface tension is measured.

It will be shown that the effect of the boundary conditions is to reweight
flux configurations
according to the $N$-ality of the flux through an interface.
With $z_1=e^{2\pi i /N}$,
configurations with $N$-ality $n$ get a factor of $z_n = z_1^n$.
Thus the partition function with interface boundary
conditions is the sum over $n$ of sets of configurations with $N$-ality $n$
and with an additional factor of $z_n$ in the weight.
This
is unlikely to be relevant to a physical situation. I do not know how
this reweighting would arise in a natural way.

When matter is added, the situation changes significantly. Matter fields in the
fundamental representation explicitly break the global $Z(N)$ symmetry.  The
$N$ line states are no longer degenerate. Above the critical temperature, there
could be metastable states of the line variables separated by
walls\cite{r8}.
However, it follows from the matter-free analysis that there are no physical
metastable states or walls in the presence of matter. A physical effect of the
matter is to lower the free energy. This comes about through an increase in the
entropy when new configurations with sources are allowed.

In this paper, I will use a very simple $Z(2)$ model to discuss these issues
from first principles and in detail. The more interesting $SU(N)$ and
$Z(N)$ gauge theories can be handled with the same approach, but they  require
more theoretical elaboration. I will cover those in another paper.

While calculations in terms of the
physical flux variables and in terms of the unphysical lines lead to the same
result, the associated interpretations are completely different.
Interfaces in the line variables may be a
convenient device for making approximate calculations of physical quantities in
terms of unphysical variables. However, one should be cautious with heuristic
arguments that rely upon the physical reality of the interfaces.
In terms of physical variables, there are no interfaces or metastable states.

Section II covers gauge fields without matter. It is shown that there are no
interfaces in the physical variables associated with the interfaces in the
lines. Boundary conditions designed to detect the contribution from line
interfaces are applied in cylinder geometry. When the problem is expressed in
the physical variables, it becomes clear that there are no associated physical
interfaces. The associated physical effect is the ``conservation'' $mod$~$N$ of
the total $N$-ality of the flux along the cylinder axis, which is perpendicular
to the line interface. The $Z(2)$ model is used to illustrate these points.
Section III introduces matter of nontrivial $N$-ality. First, there is a short
argument that because there are no physically distinct high-temperature phases
of the Yang-Mills field, there are also no metastable states in the presence
of matter. Matter simply lowers the free energy. It is shown that this
can be understood in terms of the physical variables as an increase in entropy.
Section IV is a brief conclusion.

\section{No physical walls in hot Yang-Mills theory}

This section treats gauge theory without matter fields. The Hamiltonian
formulation in the $A_0=0$ gauge is used. This is the most physical formulation
since it is closest to the elementary language of Hamiltonian  quantum
mechanics. A lattice cutoff is used. The physical degrees of freedom are gauge
group elements on links of the spatial lattice. One basis of states specifies
a definite value of the group element on each link. The other
convenient basis specifies that the amplitude to be at different points
in the group is given by a matrix element from an irreducible
representation. This is flux on links. Either is a description of physical
variables. Physical states are configurations of physical variables with the
additional condition that the Gauss law constraints are satisfied.
To enforce these constraints, the
unphysical variables are introduced in the Lagrangian formulation. These are
the group elements on links in the fourth direction or the fourth component of
the gauge field.

If the partition function is written as a sum over physical states, then the
unphysical variables have no work to do and need not even appear. Clearly,
interfaces in the unphysical variables are not physical per se. So the question
is: do they reflect the existence of interfaces in the physical
variables?

One possibility is that interfaces in the Wilson lines are associated with gaps
in the percolating flux.
There could be large islands of flux that
have no flux links to the rest of the lattice. However, such an island can be
connected to the rest of the flux network with the local addition of a finite
number of links.  For large islands, there are many ways to do this. Thus, the
probability of a large island appearing is very small---much smaller than the
probability for a wall to appear in the line variables. This follows from the
fact that a
local, finite-energy
change in the line variables does not destroy the wall, it only moves its
position a little. This conclusion can also be deduced from the calculations
below.

Previous work on the high-temperature, homogeneous phases of the
lines\cite{me,r2.1,r2.2} makes the existence of physical interfaces seem very
unlikely. It was shown that the physical configurations
associated with the $N$ pure phases of the Wilson lines at high temperature are
the same. Thus, in terms of physical variables, there is a single
high-temperature ensemble. There are not $N$ physical phases that could be
separated by walls.

So there does not appear to be any reason why there should be interfaces in the
physical variables. In particular, they are not required by symmetry as in the
case of the unphysical variables. The walls in the line variables do not imply
the existence of walls in the physical variables. The following discussion
reveals the property of physical flux that is associated with the unphysical
interfaces in the lines. It is the ``conservation'' $mod$ $N$ of flux
along the spatial direction perpendicular to the interface.

The following discussion deals mainly with the $Z(2)$ flux model. This model
is equivalent to the $Z(2)$ spin Ising model\cite{me}.
It is also an approximation to $Z(2)$ gauge theory.
Thus, the discussion and conclusions also apply to $Z(2)$ gauge theory.

The gauge and flux models are related
through the basis states and because of the limited role of the Hamiltonian at
high temperature. In the Hamiltonian formulation of the 3+1-dimensional gauge
theory, a basis for states is flux on links. More precisely, on each link, an
irreducible representation matrix element for the group is specified. The
allowed configurations are restricted to those that satisfy the Gauss law
constraints. In the case of $Z(2)$, the flux is either zero or one
on each link. The constraints are that each site must have an even number of
links with ones. These are exactly the configurations of the flux model.

In the flux model, $\theta(l)=1$ if there is flux on link $l$,
and it is 0 if there is not. The energy of a link of flux is $\sigma$.
Configurations are weighted with a Boltzmann factor
\begin{equation}
 e^{-H/T} \; \; \text{  with  } \; \; H = \sum_l \sigma \theta(l) . \label{e60}
\end{equation}
In a gauge theory, the Hamiltonian comes from the transfer matrix. There is a
factor of this form\cite{r15} and a factor from the spatial plaquette term in
the Lagrangian. The plaquette factor is diagonal in the basis where the links
have definite group elements, but it is not diagonal in the flux basis. Thus it
could complicate the discussion that follows. However, it is sufficient to
consider the high temperature limit. This is the region where the Wilson
lines are ordered. It is the large $\beta$ limit of the spin system. It is the
region where the interfaces in the lines are sharp. In a theory, such
as the $Z(2)$ case, with discrete, finite states, the Hamiltonian is a
perturbation at high temperature. Thus, the flux weighting of (\ref{e60})
plays no significant role in the discussion. That does not change if a
plaquette
term is added to $H$. Thus, the discussion in this paper is sufficient for
high-temperature, $Z(2)$, gauge theory.

It is important that there
is a discrete and finite set of flux states for each link. It might
appear that a new approach will be needed if there is an infinite number of
irreducible representations as there is for any continuous group. However,
it turns out that the role that is played by flux in the $Z(2)$ case will
be played by $N$-ality for the $SU(N)$ gauge groups. Quite a bit more machinery
is required, but the basic structure of the development is not changed.

The analysis begins with the $Z(2)$ Ising flux model used in previous
work\cite{me}.
In this model, space is a three-dimensional cubic lattice.
There are variables $\theta$ on links that can have the values 1 or 0 to
indicate the presence or absence of flux.
These are the physical variables.
A configuration is specified by the function $\theta(l)$. States with definite
flux $|\theta\rangle$ are labeled by that function.
A general state has a wave
functional that gives the amplitude for the system to be found in the various
basis states of definite flux.
The energy of a link with flux is
$\sigma$. The weight for a configuration is
\begin{equation}
  \langle \theta | e^{-H/T} | \theta \rangle =
 e^{- \case{1}{T} \sum_{l} \sigma \theta(l)} =
            \prod_{l} e^{- \case{1}{T} \sigma \theta(l)}   . \label{e101}
\end{equation}
The sum over configurations is restricted to those in which the number of links
at a site that have flux is even. Let the collection of all such
configurations be $C'$. It is a subset of the unrestricted collection
of all configurations $C$. The partition function is
\begin{equation}
 Z = Tr'[e^{-H/T}] \equiv \sum_{C'}
                           \langle \theta | e^{-H/T} | \theta \rangle
      = \sum_{C'} e^{- \case{1}{T} \sum_{l} \sigma \theta(l)} .
\end{equation}

The equivalence of this flux model to the Ising model
results from using site variables to enforce the restriction
on configurations, and then doing the $\theta$ sums.
The site variables are the Ising spins.
First, consider the sum of $\theta(l)$
over the $2d=6$ $l$'s contained in the set $I(i)$ of links with endpoint $i$:
\begin{equation}
  \Sigma(i) \equiv \sum_{l \in I(i)} \theta(l)  .
\end{equation}
To force this to be even at each site, introduce the site variables $s(i)$
that take the values $\pm 1$. The factor
\begin{equation}
 \case{1}{2} \sum_{s(i)=\pm 1} s(i)^{\Sigma(i)}            \label{e102}
\end{equation}
has the desired effect. A factor like this is introduced into the
partition function sum for each site $i$. The spins $s(i)$ are the same as the
Wilson lines of the $Z(2)$ gauge theory. This model is equivalent to the Ising
model for the spins $s(i)$ with the Ising $\beta$ related to the gauge $T$ by
$e^{-\sigma/T}=\tanh \beta$.
Note that I never refer to the Ising model $1/\beta$ as temperature.
Temperature always refers to the $T$ appearing in (\ref{e101}).
Large $T$ gives large $\beta$, and ordered Ising spins.
It is convenient to introduce $\mu \equiv \sigma / T$.
In the high $T$ region,
\begin{equation}
  e^{-2\beta} \sim \mu/2             \label{e80}
\end{equation}

To discuss interfaces, it helps to introduce boundary conditions that force at
least one interface in the lines to appear. Use
solid-cylindrical geometry: space is finite in the transverse $x$ and $y$
directions and very long in the
$z$ direction.
There are periodic boundary conditions in the $x$ and $y$ directions. The
transverse area in dimensionless lattice units is $A$.
Let $L$ be large and positive.
Apply the boundary conditions $s=-1$ for $z = -L/2$ and $s=1$ for $z = L/2$.
The limit of interest is $L \rightarrow \infty$ followed by
$A \rightarrow \infty$.

First review the picture in terms of the unphysical spin variables.
The high-temperature region is
the large-$\beta$, ordered phase for the Ising spins.
The interface is approximately flat and
has a Boltzmann weight $e^{-2\beta A}$.
Thus, the energy per unit area $\alpha$ is related to $\beta$
by $\alpha/T \approx 2\beta$ as $\beta \rightarrow \infty$.
With $L-1$ planes between $z = -L/2$ and $z = L/2$, there
are $L$ places to put an interface. The boundary conditions require an odd
number of them.
Since $\beta$ is large, $\alpha$ is large, and the interfaces are sparse.
The sum in the partition function can be extended to infinity to obtain
\begin{equation}
 2Z' = e^{L e^{-\alpha A/T}} - e^{-L e^{-\alpha A/T}}  .
\end{equation}
The partition function $Z$ with the same boundary conditions
at each end is
\begin{equation}
 2Z = e^{L e^{-\alpha A/T}} + e^{-L e^{-\alpha A/T}}  .
\end{equation}
The ratio is
\begin{equation}
  Z'/Z = e^{-(F'-F)/T} \cong 1 - (F'-F)/T
       = 1 - 2 e^{-2L e^{-2\beta A}} . \label{e21}
\end{equation}
The excess free energy
\begin{equation}
 F'-F = 2 T e^{-2 L e^{-2\beta A}}
\end{equation}
decays to 1 exponentially at a rate determined by the interface energy per unit
area $\alpha = 2\beta T$. This relates the activity for the wall
$e^{-2\beta A}$ to $Z'/Z$ and $F'-F$. The approximations are valid for
large $L$ and $Le^{-2\beta A} \gg 1$ .

Now consider the same situation in terms of flux variables.
In the high-temperature, deconfined phase with $\mu= \sigma/T$ small,
there is dense, percolating flux.
The flux is almost random except for the
constraint that there be an even number of links with flux at each site.
The following discussion shows how the boundary conditions, which require at
least one interface, affect the flux configurations contributing to the
partition function.

The system consists of
all the links and sites strictly between the two planes with the fixed spins.
To enforce the constraints, there is a factor of
\begin{equation}
 \frac{1}{2} \sum_s s^{\sum \theta}
\end{equation}
at each site. For a site on the $z=-L/2$ boundary, this becomes
$(-1)^{\theta}$.
The longitudinal links that end on
boundary sites are not constrained on the boundary end because the spin is
fixed there. Of course, the
constraints at the sites on the other ends of those links must still be
satisfied. The boundary links can be freely chosen to satisfy the constraints
on the first layer of spins at each end. On the positive $z$ end,
nothing else happens. However, on the negative $z$ end, there is a factor of
$-1$ for each link with flux coming into the system from the boundary.

{}From the local Gauss law constraints, it follows that
the total flux $\sum_{xy} \theta_z(x,y,z)$ $mod$ 2 on a transverse slice of
longitudinal links is independent of $z$.
This will be referred to as flux ``conservation''. Thus each configuration can
be described as
even or odd. The effect of the boundary conditions is to weight the odd
configurations with and extra factor of $-1$. The partition function sum
\begin{equation}
 Z = \sum_{even} e^{-H/T} + \sum_{odd} e^{-H/T}
\end{equation}
is replaced by
\begin{equation}
 Z' = \sum_{even} e^{-H/T} - \sum_{odd} e^{-H/T}   .
\end{equation}
This gives
\begin{equation}
  Z'/Z =  (Z_e - Z_o)/(Z_e + Z_o) = 1-2e^{-(F_o-F_e)/T} .  \label{e20}
\end{equation}
The last step is correct for large $(F_o-F_e)$.
I will show in a moment that this free energy difference is
proportional to the
length $L$, so that (\ref{e20}) is consistent with (\ref{e21}).
Thus, the free energy difference per unit length of odd flux verses even flux
is the quantity of interest.

The next part of the discussion shows the direct connection between the spin
interfaces and the conservation of flux.
Consider the structure of the partition function.
The role of the spins is to enforce the local constraints on flux.
In the calculation leading to (\ref{e21}),
the spins are constant on transverse planes. Thus, they are enforcing the
weaker constraint of flux conservation. The details follow.

Let the constant value of
the spins at position $i$ along the longitudinal axis be $S(i)$.
In this situation,
the flux on transverse links is unconstrained. In the partition function,
the transverse flux sum gives the same
factor on each plane. That factor is raised to the power $L-1$.
This is an uninteresting, overall factor.
Another ingredient in the partition function is the flux sum on
a single slice of longitudinal links.
There are $A$ longitudinal links into a transverse plane.
The sum becomes a product of A independent sums and gives
\begin{equation}
  \sum_{\theta} e^{-\mu \sum \theta} [S(i) S(i+1)]^{\sum \theta}
               =  [ 1 + e^{-\mu} S(i) S(i+1)]^A
   =  a + b S(i) S(i+1)
\end{equation}
with
\begin{equation}
 a = \frac{1}{2} [ (1 + e^{-\mu})^A + (1 - e^{-\mu})^A]
       \;\; \text{ and } \;\;
 b = \frac{1}{2} [ (1 + e^{-\mu})^A - (1 - e^{-\mu})^A] .
\end{equation}

Now consider the calculation of $Z'$.
Let $\Theta(i) \equiv \sum_j \theta(j,i)$ be the sum of all the longitudinal
flux on a slice at position $i$.
\begin{eqnarray}
 Z' & = & \sum_{\theta} \sum_S (-1)^{\Theta(1)}
         e^{-\mu \Theta(1)} S(1)^{\Theta(1)+\Theta(2)}
         e^{-\mu \Theta(2)} S(2)^{\Theta(2)+\Theta(3)}  \ldots \\
    & &    e^{-\mu \Theta(N)} S(N)^{\Theta(N)+\Theta(N+1)}
         e^{-\mu \Theta(N+1)} .
\end{eqnarray}
First do the $\theta$ sums to obtain
\begin{equation}
 Z' = \sum_S [a - b S(1)] [a + b S(1) S(2)] \ldots [a + b S(N-1) S(N)]
                                               [a + b S(N)] .
\end{equation}
This brings us back to a one-dimensional Ising model. The spin sums can be
done. The results are
\begin{equation}
 Z' = a^L - b^L      \label{e30}
\end{equation}
and
\begin{equation}
 Z'/Z = 1 - 2 e^{-L \ln (a/b)} . \label{e22}
\end{equation}
For small $\mu$, $a/b \sim 1+2(\frac{\mu}{2})^A$. Comparing (\ref{e22}) with
(\ref{e20}) gives
\begin{equation}
 Z'/Z = 1 - 2 e^{-2L(\frac{\mu}{2})^A}
\end{equation}
and
\begin{equation}
  (F_o-F_e)/T = 2L(\frac{\mu}{2})^A  .
\end{equation}
Since $\mu$ and $\beta$ are related by (\ref{e80})
this is equivalent to (\ref{e21}).

This shows that the interfaces in the unphysical
spin variables are associated with flux conservation i.e.\ the
total flux $mod$ 2 on a transverse slice of links is independent of
longitudinal position.

With that understood, one can redo the calculation of $Z'$ by a much easier
method using flux variables. The partition function $Z'$ is the
difference of two contributions. It is the sum of all the even configurations
minus the sum of all the odd configurations. The weight for one slice of even
longitudinal flux is
\begin{equation}
 z_e = \sum_{\theta} \frac{1}{2} \sum_s e^{-\mu \Sigma \theta}
             s^{\Sigma \theta} = a   .
\end{equation}
For odd longitudinal flux, it is
\begin{equation}
 z_o = \sum_{\theta} \frac{1}{2} \sum_s e^{-\mu \Sigma \theta}
             s^{\Sigma \theta + 1} = b .
\end{equation}
For small $\mu$, the configurations of flux on longitudinal links are almost
independent from slice to slice except that the total flux $mod$ 2 is the same.
Thus
\begin{equation}
 Z' = Z_e - Z_o = z_e^L - z_o^L = a^L - b^L
\end{equation}
as in (\ref{e30}).
This is almost zero because the weights for even and odd configurations are
almost the same when $\mu$ is small.

This calculation of $Z'$ is simple and direct in the physical flux variables.
There is no reference to interfaces. The result is the same as that from the
spin calculation, but the associated physical picture is completely different.
In flux, it is the Gauss law constraints rather than interfaces that are
important.

This completes the main discussion of the matter-free case. Before considering
the introduction of matter in the next section, let us examine the
approximations that have been used.
The main approximation has been to assert that after summing over
flux configurations on transverse links, the configurations on longitudinal
links from slice to slice are independent except for the
restriction that they have the same total flux $mod$ 2.
This is equivalent to the statement that at
small $\mu$ and large $\beta$, the spins on a transverse slice are ordered.

To examine the approximation, fix a configuration of flux on
longitudinal links subject only to the
constraint of flux conservation. Consider a plane of
transverse links separating two slices of longitudinal links.
In the large $T$, $\mu \rightarrow 0$ limit, the allowed configurations get
equal weight. The density of flux is high. Without the constraints, it would be
$0.5$ per link. The constraints say that only configurations with an even
number
of links with flux at each site are allowed. If there were no longitudinal
flux, then that constraint would apply to the transverse flux on the plane.
The effect of the longitudinal flux is to require odd transverse flux at some
sites. Since longitudinal flux is conserved,
there are an even number of such sites on each slice. The
crucial fact is that the number of allowed configurations of transverse flux is
independent of the number of pairs of odd sites. Given that, it is clear that
the sum over all transverse flux configurations is independent of the
longitudinal configuration. This is the approximation that was used above.

The problem is reduced to showing that the number of allowed configurations
is independent of the number of pairs of odd sites. Consider two configurations
of pairs of odd sites $c$ and $c'$. The corresponding sets of allowed flux are
$C$ and $C'$. I need to show that the numbers of configurations in $C$ and $C'$
are the same. In spin variables, this is just the statement that n-point
functions of spins are position independent at infinite $\beta$. In the flux
picture, the result can be established by producing a one-to-one mapping
between $C$ and $C'$. Without loss of generality, assume that $c$ and $c'$
differ by $c'$ having a single extra pair of odd sites relative to $c$. Connect
those two sites with some path of links. Reversing the flux $0
\leftrightarrow 1$ on the links of the path gives the desired map.

The conclusion of this section is that with the cylinder
geometry and boundary conditions that reveal the presence of walls
in spin variables, there are no walls in physical flux variables.
The spin walls are associated with flux conservation along the cylinder axis.
The ratio $Z'/Z$ of partition functions with odd and even boundary conditions
can be used to extract the surface energy of a wall. In the flux picture, it is
a measure the difference in free energy per unit
length between configurations with even flux and those with odd flux on the
longitudinal links.

However, the universe does not have this artificial geometry and boundary
conditions. For this free energy difference
to be physically relevant, it would require that a factor of
$-1$ be associated with a closed surface when the flux through it is odd. This
was contrived in the discussion above through the choice of geometry and
boundary conditions. I do not see how it could arise naturally. Thus, the spin
interface tension has a physical counterpart but it is not physically
relevant.

\section{Effects of matter}

With the introduction of matter carrying an irreducible representation
of $SU(N)$ or $Z(N)$ with
nontrivial $N$-ality,
the global $Z(N)$ symmetry is explicitly broken.
This lifts the degeneracy of the multiple, high-$T$ phases of the line
variables.
When the effect from the matter is small, there will be metastable states.
Such states and their possible astrophysical consequences have been
discussed\cite{r8}. The important effect of the matter is that it breaks the
global symmetry.
This can be studied in a simpler situation where
the matter is represented by an external field $h$ linearly coupled to the
Wilson lines: $h^* L + h L^* $.

In a general situation with $Z(2)$ global symmetry, the constraint effective
potential with $h \neq 0$ is a double well with one side lower than the other.
As $h \rightarrow 0$, this becomes the symmetric double well with degeneracy.
Thus the existence of the metastable state is linked to the existence of
degeneracy. This is the generic situation away from the critical point in the
two-phase region.

However, without matter, there is a single, physical, high-temperature phase.
Thus, there is no degeneracy that could be connected to multiple states,
some metastable, when symmetry breaking is present. I conclude that
in the physical flux picture, degeneracy, interfaces, and
metastable states are all absent.

Nevertheless, matter does affect the free energy. It is interesting to
describe this effect entirely in the flux picture. The idea is to see what is
happening to the flux configurations when one side of the effective potential
for the lines is lowered. Lowering the free energy
means increasing the partition function. I will show that this increase in $Z$
comes from an increase in the number of allowed flux configurations in the
presence of matter i.e.\ an entropy increase.

It is sufficient for this discussion to consider the case of heavy,
dilute matter. It follows from the work in Ref.\ top
\cite{r12} that this is
well approximated by a small external field.
Thus, in the $Z(2)$ flux model, the symmetry breaking effect of matter will be
represented by a weak external field $h$ coupled linearly to the line or spin
variables.

In the high-temperature, two-phase region, the free energy is lowered. When
the partition function is expressed in
terms of flux variables, the effect of $h$ is to allow odd sites with a
density controlled by $h$. I will show that this increases the number of
allowed configurations, increases $Z$, and decreases $F$.

The extra factor that comes from the introduction of matter (as represented by
the external field) is
\begin{equation}
 e^{ \sum_i h s(i)} = \prod_i e^{h s(i)} =\prod_i (\cosh h + s(i) \sinh h)
\end{equation}
The sums on the spin variables give the constraints on flux. For each
site, there is a factor of
\begin{equation}
   \delta_E \cosh h +  \delta_O \sinh h  .
\end{equation}
The shorthand $\delta_{(E,O)}$ stands for the restriction that the flux at the
site be (even, odd). With $h$ small, most sites have even flux, but there is a
small density of sites with odd flux.
The sites with even flux have a slightly reduced weight $\cosh h$ vs.\ 1,
while sites with odd flux are now allowed with a small weight $\sinh h$.
I will show that
the latter effect is more important so that $F$ decreases while $Z$ increases.

A little notation is convenient.
Let $Z(i_1, \ldots, i_n)$ be the partition function from summing over
configurations with odd flux at the sites $i_1, \ldots, i_n$ and even flux at
all others. In finite volume $V$, $n$ must be even. Further, define
$Z(i_1, \ldots, i_n)$ so that it vanishes if $n$ is odd or if any two $i$'s
are equal. Let $Z(h)$ be the partition function with the
external field $h$.
\begin{eqnarray}
 Z(h) & = & \sum_{\theta} e^{- \frac{1}{T} \sum_l \sigma \theta(l)}
        \prod_i  [ \cosh h \delta_E + \sinh h \delta_O ]  \\
      & = & Z(0) (\cosh h)^V [ 1 + \sum_{n} (\tanh h)^n \frac{1}{n!}
             \sum_{i_1, \ldots, i_n} Z(i_1, \ldots, i_n)/Z(0)   \label{e5}
\end{eqnarray}
Since only even $n$'s are present, $Z(h)$ is even in $h\rightarrow-h$.

It is useful to extend the definition of $Z$ to odd $n$. Factorization
properties of $Z$ are also needed.
It will be shown that there is a function
$\tilde{Z}(i_1,\ldots,i_n)$, defined for even and odd $n$ with the following
properties. For even $n$, $\tilde{Z}(i_1,\ldots,i_n)=Z(i_1,\ldots,i_n)$.
To describe the factorization, let $P=\{p_1,p_2,\ldots\}$ be a partition of
$\{i_1,\ldots,i_n\}$.
When the clusters of points are widely separated,
\begin{equation}
  \tilde{Z}(i_1,\dots,i_n) \rightarrow \prod_i \tilde{Z}(i\in p_1) \label{e52}
{}.
\end{equation}
For example,
\begin{equation}
  \tilde{Z}(i_1,i_2) = Z(i_1,i_2)
       \rightarrow \tilde{Z}(i_1) \tilde{Z}(i_2) \; \; {\text{ as }} \; \;
        |i_1 - i_2| \rightarrow \infty
\end{equation}
and
\begin{equation}
  \tilde{Z}(i_1,i_2,i_3) \rightarrow \tilde{Z}(i_1,i_2) \tilde{Z}(i_3)
       \;  {\text{ for }} \;  |i_3 - i_1| \rightarrow \infty \;
         {\text{ with }} \; i_1 \; {\text{ and }} \; i_2 \; {\text{ fixed}}.
\end{equation}
Note that for odd $n$, $\tilde{Z}$ is not a finite-volume partition function
with an odd number of odd-flux sites. $\tilde{Z}(h)$ is defined by replacing
$Z$ with $\tilde{Z}$ in (\ref{e5}).
The original $Z(h)$ can be recovered via
\begin{equation}
 Z(h) = \frac{1}{2} [ \tilde{Z}(h) + \tilde{Z}(-h) ]  .      \label{e40}
\end{equation}
It is a bit more convenient to deal with
\begin{equation}
 \tilde{Z}' =  1 + \sum_{n} (\tanh h)^n \frac{1}{n!}
             \sum_{i_1, \ldots, i_n} \tilde{Z}(i_1, \ldots, i_n)/\tilde{Z}(0) .
\end{equation}

Now introduce $W(h)$ by
\begin{equation}
 \tilde{Z}(h) = e^{W(h)}                        \label{e41}
\end{equation}
and write
\begin{equation}
 W(h) = \sum_{n=1}^{\infty} (\tanh h)^n \frac{1}{n!}
      \sum_{i_1, \ldots, i_n} W(i_1, \ldots, i_n) .   \label{e1}
\end{equation}

$W(i_1, \ldots, i_n)$ is defined recursively from $\tilde{Z}(i_1,\ldots,i_n)$
as usual\cite{r13,r14}.
If $\tilde{Z}(i_1,\ldots,i_n)$ factorizes for large separation, then
$W(i_1,\ldots,i_n)$ goes to zero if any separation between $i$'s is
large. Thus, the sum on the $i$'s in (\ref{e1}) is proportional to the spatial
volume $V$.
The quantity of interest is the coefficient of $V$ in the $h \rightarrow 0$
limit. This comes from $w \equiv W(i)$.
Combining (\ref{e5}), (\ref{e40}), (\ref{e41}), and (\ref{e1}) gives
\begin{equation}
 Z(h) = Z(0) (\cosh h)^V \frac{1}{2} [e^{V|hw|} + e^{-V|hw|}]
\end{equation}
for $V$ large and $h$ small.
In this limit, the $e^{V|hw|}$ term
dominates. Thus as $h$ increases from zero, there is an
increase in $Z$ and a decrease
in the free energy. It is clear from the discussion
that this is a direct consequence of the increase in entropy that comes from
including some configurations with
odd-flux sites. This effect dominates the middle factor $(\cosh h)^V$,
which reflects the
change in the weight of the configurations with even flux at every site.
This completes our discussion of the behavior of the system in terms of the
physical variables. However, the conclusions are based on properties of $Z$ and
$\tilde{Z}$ that must now be established.

\subsection{Properties of $Z(i_1, \ldots, i_n)$}
It must be shown that $Z$ has factorization properties which allow
$\tilde{Z}$ to be defined and to factorize according to (\ref{e52}).
First recall that $T$ is large and the flux is
dense. I will use an expansion in
$y(l) \equiv e^{-\frac{1}{T}\sigma \theta (l)} - 1$, which
is small when $T$ is large.
Let $C(i_1,\ldots,i_n)$ be
the set of flux configurations with odd flux at the indicated sites and even
flux at all others. The number of configurations in $C$ is denoted by $\#C$.
Clearly, $n$ must be even.
\begin{eqnarray}
   Z(i_1, \ldots, i_n) & = & \sum_{C(i_1,\ldots,i_n)} \prod_{l}
         e^{-\frac{1}{T}\sigma \theta (l)}  \\
 & = & \sum \prod [1+y(l)] \\
 & = & \sum [1+ \sum_{l_1} y(l_1) + \sum_{\{l_1,l_2\}} y(l_1) y(l_2) + \ldots]
                    \label{e50} \\
 & = & \sum [ 1 + \sum_m \frac{1}{m!} \sum_{l_1 \ldots, l_m | l_i \neq l_j}
                                    y(l_1) \ldots y(l_m)]  \label{e51}
\end{eqnarray}
An $m$-fold $l$ sum in (\ref{e50}) is over all subsets of the links with
$m$ elements.
In (\ref{e51}), the $l$-sums are restricted only by the condition that no two
links can be the same. Introduce the notation $z(j)$ and $w(j)$ for the
generating functions of the $y(l)$'s.
\begin{eqnarray}
 z(j) & = & e^{w(j)} = \frac{1}{\# C}  \sum_C [ 1 + \sum_m \frac{1}{m!}
 \sum_{l_1, \ldots, l_m}'' y(l_1) \ldots y(l_m) j(l_1) \ldots j(l_m) ]
                                                           \label{e91} \\
  & \equiv & [ 1 + \sum_m \frac{1}{m!}
    \sum_{l_1, \ldots, l_m} z(l_1, \ldots, l_m) j(l_1) \ldots j(l_m) ]
                                                                 \label{e90}
\end{eqnarray}
and
\begin{equation}
 w(j) =    \sum_{m=1}^{\infty} \frac{1}{m!}
 \sum_{l_1, \ldots, l_m} w(l_1, \ldots, l_m) j(l_1) \ldots j(l_m)   .
\label{e3}
\end{equation}
The dependence of $z$'s and $w$'s on the $i$'s has not been shown here.
Equation (\ref{e90}) is a definition of the $z(l_1,\ldots,l_m)$. The $l_i$ sums
are unrestricted in (\ref{e90}) so $z(l_1,\ldots,l_m)$ vanishes if any $l$'s
coincide.

The key technical result is a factorization property of $z(l_1, \dots, l_m)$.
Let $R$ be
a partition of the links $\{l_1, \dots, l_m\}$ into subsets $r_i$.
Whenever the links of each $r_i$ are well separated from those of
all other $r_j$, then
\begin{equation}
 z(\{l \in R \}) = \prod_i z(\{l \in r_i \})  .   \label{e2}
\end{equation}
This will be proven below.

Given (\ref{e2}), the usual combinatoric arguments lead to the conclusion
that $w(l_1, \dots, l_n)$ falls to zero rapidly if any link separation becomes
large. This means that the sums in (\ref{e3}) can be considered as a sum on all
possible clusters of links followed by a sum that moves the cluster to all
possible positions. Since every link that is included brings a factor of
$y \sim 1/T$, only small clusters contribute at high temperature.

Consider the case in which
the odd sites are partitioned into groups $p_i$ that are widely separated. The
objects of interest are
\begin{equation}
 w(j=1) =
  \sum_m \frac{1}{m!}
 \sum_{l_1, \ldots, l_m}
         w(i\in p_1, i\in p_2,\ldots; l_1,\ldots, l_m)
\end{equation}
and
\begin{eqnarray}
  \ln \frac{Z(i_1,\dots,i_n)}{Z(0)} & = &
  \sum_m \frac{1}{m!}
 \sum_{l_1 \ldots, l_m}
    [ w(i\in p_1, i\in p_2,\ldots; l_1,\ldots, l_m) -w(l_1,\ldots, l_m)] \\
      & = & \bar{w}(p_1) + \bar{w}(p_2) +\ldots + \bar{w}(rest)  \label{e53} .
\end{eqnarray}
$\bar{w}(p_i)$ is the contribution from the part of the $l$ sums
where the cluster of links $\{l_1,\ldots,l_m\}$ is around the group of sites
$p_i$. $\bar{w}(rest)$ is from the terms in the $l$ sums where the cluster of
links is far from any odd site.
When the cluster of links $\{l_1,\ldots,l_m\}$ is far from an odd site $i$,
then the flux configurations around the cluster, which determine the value of
$w$ via (\ref{e91})--(\ref{e3}), are independent of the position $i$.
This is because $C(i,j,\ldots,k)$ can be mapped to $C(i',j,\dots,k)$ without
changing the configurations away from $i$ and near $j,\ldots,k$. This can also
be shown using the technique introduced at the end of this section.
Thus, the last term $\bar{w}(rest)$ is independent of $i$.

So, $Z(i_1,\ldots,i_n)$ is a product of factors---one for each cluster $p_j$
of odd sites. The number of sites in a cluster can be even or odd. These
factors can be used to define $\tilde{Z}(i_1,\ldots,i_n)$ for odd $n$. So
defined, the $\tilde{Z}(i_1,\ldots,i_n)$ have the required properties.

Equation (\ref{e53}) shows that
\begin{equation}
 W(i)= \bar{w}(i)     .
\end{equation}
It is straightforward to calculate $\bar{w}(i)$ in low orders of the
high-temperature
expansion. The first nonzero term comes at order $T^{-6}$ when the site $i$ has
contributing $y$'s on all the links attached to it. The result is
\begin{eqnarray}
    e^{W(i)} & = & 1 - \frac{1}{32} (1 - e^{-\sigma / T})^6 \\
             & = & 1- 2 e^{-12 \beta}        \label{e54}                \\
             & = & \langle s(i) \rangle
                      \; \; ({\text {to this leading order}}).  \label{e55}
\end{eqnarray}
The the right hand sides of (\ref{e54}) and (\ref{e55}) could also be
obtained from a simple calculation in terms of spin variables.
This shows that the flux calculation produces the same result as the spin
calculation.

Finally, the factorization of
$z$ in its $l$ variables shown in (\ref{e2}) must be derived.
This requires several steps and a little notation.
The configurations have odd flux at sites $i_1,\dots,i_n$ only.
This is understood and not explicitly displayed in the following discussion.
Subsets of links are indicated by $X$, $X_1$, etc.
A specific configuration of $\theta(l)$'s, $l \in X$ is denoted by $\Theta$.
The set $C(\Theta_1,\Theta_2,\ldots)$ is all allowed configurations
of $\theta$'s with given, fixed subconfigurations on the specified subsets of
links and with odd sites at $i_1,\dots,i_n$.

The derivation establishes the following assertions:
\begin{enumerate}
 \item There is a one-to-one map $C(\Theta) \leftrightarrow C(\Theta')$.
 \item Thus, $\#C(\Theta) = \#C(\Theta')$.
 \item If two subsets of links $X_1$ and $X_2$ are separated, then a
configuration of $\theta$'s on $X_2$ is allowed or not independent of the
configuration of the $\theta$'s on $X_1$.
 \item $\#C(\Theta_1,\Theta_2) = \#C(\Theta_1,\Theta_2')$
 \item There is a factorization
\begin{equation}
 \langle f_1(\Theta _1) f_2(\Theta _2) \rangle _0 =
  \langle f_1(\Theta _1)\rangle _0 \langle f_2(\Theta _2) \rangle _0  .
                     \label{e56}
\end{equation}
\end{enumerate}
The subscript $0$ indicates an ensemble in which all the contributing
configurations are equally weighted---the $T=\infty$ limit.
The relation (\ref{e2}) is a special case of (\ref{e56}).

Here is the reasoning that establishes each of the assertions.
\begin{enumerate}
 \item Consider a configuration $c_0$ in which the subconfiguration on $X$ is
$\Theta$ and another configuration $c'_0$ with $\Theta'$ on $X$.
For $c_0$, let $Y$ be the set of links
on which $\theta(l)=1$, and for $c_0'$, the corresponding set is $Y'$.
The map is defined as a reversal ($0 \leftrightarrow 1$) of the $\theta$'s on
links in $U = (Y \cup Y') - Y \cap Y'$. This is an invertible map on and onto
$C(i_1,\ldots,i_n)$ that sends $c_0 \rightarrow c_0'$.
To see that the same map, with $c_0$ and $c_0'$ unchanged, sends any
configuration in $C(\Theta)$ to one in $C(\Theta')$, consider links $l \in X$.
If $\theta(l)=1$ in $\Theta$ and $\Theta'$, then
$l \in Y \cap Y'$ and $l \not\in U$, and $\theta(l)$ does not change.
If $\theta(l)=0$ in $\Theta$ and $\Theta'$, then
$l \not\in Y \cup Y'$ and $l \not\in U$, and $\theta(l)$ does not
change.
If $\theta(l)=1$ in $\Theta$ and $\theta(l)=0$ in $\Theta'$, then
$l \in Y$ and $l \not\in Y'$ and $l \in Z$, and $\theta(l)$ is reversed,
as it should. Clearly, the fourth case works like this one.
 \item This is an immediate consequence of the fact that the map is one-to-one
and onto.
 \item Let $\bar X$ be a cube of links which contains $X_2$. Let $R$ be
a larger cube of links containing $\bar X$. The region outside $R$,
which contains $X_1$ is
$V_0$, and the region in $R$ but not in $\bar X$ is $I$. For any
configuration on $V_0 \cup \bar X$, I can change links in $I$ so that
there is either no path of $\theta=1$ links or one path of $\theta=1$ links
that
connects $\bar X$ and $V_0$. This depends on whether the number of odd sites in
$\bar X$ is even or odd. Now change the configuration on $X_2$
to any other allowed configuration. With appropriate changes in $I$ and no
changes in $V_0$, it can be connected to $V_0$.
Thus, an allowed configuration on $X_2$ is allowed independent of the
configuration of $\theta$'s well separated from it and, in particular, those in
$X_1$.
 \item With $X_1$ and $X_2$ well separated, take $X = X_1 \cup X_2$.
The previous steps combine to show that there is a one-to-one map
$C(\Theta_1,\Theta_2) \leftrightarrow C(\Theta_1,\Theta_2')$. Thus,
$\#C(\Theta_1,\Theta_2) = \#C(\Theta_1,\Theta_2')$.
 \item Consider
\begin{eqnarray}
 \langle f_1(\Theta_1) f_2(\Theta_2) \rangle_0 & = &
   \frac{1}{\#C} \sum_{C} f_1(\Theta_1) f_2(\Theta_2) \\
& = & \frac{1}{\#C} \sum_{\Theta_1} \sum_{\Theta_2} \sum_{C(\Theta_1,\Theta_2)}
f_1(\Theta_1) f_2(\Theta_2) \\
& = & \frac{1}{\#C} \sum_{\Theta_1} f_1(\Theta_1) \sum_{\Theta_2}
f_2(\Theta_2)
\sum_{C(\Theta_1,\Theta_2)} \\
& = & \{ \frac{1}{\#C}  [\sum_{\Theta_1}]
[\sum_{\Theta_2}] [\sum_{C(\Theta_1,\Theta_2)}] \}
  \langle f_1(\Theta _1)\rangle _0 \langle f_2(\Theta _2) \rangle _0 .
                                                             \label{e93}
\end{eqnarray}
The first factor $\{\ldots\}$ in (\ref{e93}) is one, and the relation is
established.
\end{enumerate}

In this section, I have shown that the effect of fundamental
representation matter can be understood in physical variables. The usual
picture in terms of unphysical variables is that the $Z(2)$ symmetry is broken,
and the double well of the effective potential becomes asymmetric with one side
lowered in free energy and the other raised to a metastable state. In terms of
physical variables, there is no double structure with a metastable state. The
free energy is simply lowered. This results from an increase
in the entropy that comes from allowing some new configurations where the
gauge field does not transform as a singlet at the sites with matter.

\section{Conclusion}

In terms of the Wilson line variables, the high temperature phase of a gauge
theory is associated with broken symmetry. This is a multiphase region. Phase
separation and domain walls are possible. However, in a Hamiltonian
formulation, the lines are unphysical variables so it is interesting to
formulate the results in terms of physical variables and physical states.
I have done this for the $Z(2)$ case. In terms of the physical variables, there
is only one high temperature phase. There are no interfaces. The interface
tension of the lines has a physical counterpart. It arises when
configurations with odd flux are given an extra factor of $-1$ in the sum for
the partition function. It is hard to see what physical situation might make
this modification relevant.

When matter that breaks the center symmetry is added, there is a
lowering of the free energy of one of the high temperature phases of the lines.
The other phases
are then metastable. In the physical variables, there are no metastable states.
However, the free energy of the single
high-temperature phase is lowered by the matter. The origin of that is
an increase in the
entropy that follows from allowing some new configurations where the gauge
field does not transform as a singlet at the sites with matter.

\acknowledgments
This research was supported by the United States Department of Energy.


\begin{references}

\bibitem{r1}
B. Svetitsky and L. Yaffe, Nucl. Phys. {\bf B210[FS6]}, 423 (1982);
       B. Svetitsky, Phys. Rep. {\bf 132}, 1 (1986).

\bibitem{r2}
J. Boorstein and D. Kutasov, Phys. Rev. D{\bf 51}, 7111 (1995);
T. Bhattacharya, A. Gocksch, C. Korthals Altes, and R. Pisarski,
   Phys. Rev. Lett. {\bf 66}, 998 (1991);
   Nucl. Phys. B{\bf 383}, 497 (1992);
R. Pisarski, Brookhaven National Laboratory,
                     Report No.\ BNL-P-2/92, hep-ph/9302241;
A. Gocksch and R. Pisarski, Nucl. Phys. B {\bf 402}, 657 (1993).

\bibitem{me}
J. Kiskis, Phys. Rev. D{\bf 51}, 3781 (1995).

\bibitem{r2.1}
A. Smilga, Ann. Phys. {\bf 234}, 1 (1994);
A. Smilga, Report No.\ SUNY-NTG-95-34, hep-ph/9508305.

\bibitem{r2.2}
I. Kogan, Phys. Rev. D{\bf 49}, 6799 (1994);
T. Hansson, H. Nelsen, and I. Zahed, University of Stockholm,
        Nucl. Phys. B{\bf 451}, 162 (1995).

\bibitem{r2.3}
V. Belyaev, I. Kogan, G. Semenoff, and N. Weiss,
                          Phys. Lett. B {\bf 277}, 331 (1992);
W. Chen, M. Dobroliubov, and G. Semenoff,
                          Phys. Rev. D {\bf 46}, R1223 (1992).

\bibitem{r2.4}
M. Faber, O. Borisenko, and G.Zinovjev, Nucl. Phys. {\bf 444}, 563 (1995)
 and references therein.

\bibitem{r3}
J. Kogut and L. Susskind, Phys. Rev. D{\bf 11}, 395 (1975);
J. Kogut, D. Sinclair, and L. Susskind, Nucl. Phys. B{\bf 114}, 199 (1976).

\bibitem{r6}
K. Kajantie, L. Karkkainen, and K. Rummukainen,
     Nucl. Phys. B{\bf 357}, 693 (1991);
R. Brower, S. Huang, J. Potvin, C Rebbi, and J. Ross,
                  Phys. Rev. D{\bf 46}, 4736 (1992).

\bibitem{r8}
V. Dixit and M. Ogilvie, Phys. Lett. B{\bf 269}, 353 (1991);
J. Ignatius, K. Kajantie, and K. Rummukainen,
       Phys Rev. Lett. {\bf 68}, 737 (1992);
C. Korthals Altes, K. Lee, and R. Pisarski,
       Phys. Rev. Lett. {\bf 73}, 1754 (1994).

\bibitem{r15}
J. Drouffe and C. Itzykson, Phys. Rep. {\bf 38}, 133 (1978).

\bibitem{r12}
T. Blum, J. Hetrick, and D. Toussaint, Report No.\ AZPH-TH/95-22,
hep-lat/9509002.

\bibitem{r13}
For example, C. Itzykson and J.-B. Zuber, Quantum Field Theory, McGraw-Hill,
New York, 1980, p. 270.

\bibitem{r14}
Note that even though the $z$'s vanish for coinciding points, the $w$'s do not.

\end{references}
\end{document}